\documentclass[10pt,letterpaper,twocolumn]{article} %% two column, final layout

\usepackage{ol2}
\usepackage[draft]{hyperref}
\usepackage{amsmath}

\begin{document}

\twocolumn[ %% activate for two-column option

\title{Third and fifth harmonics generation by tightly focused femtosecond pulses 
at 2.2\,$\mu$m wavelength in air}

\author{Gombojav O. Ariunbold, Pavel Polynkin,$^{*}$ and Jerome V. Moloney}

\address{College of Optical Sciences, The University of Arizona, Tucson, Arizona 85721, USA
\\
$^*$Corresponding author: ppolynkin@optics.arizona.edu
}

\begin{abstract}
We report experiments on the generation of third and fifth harmonics of millijoule-level, tightly focused, 
femtosecond laser pulses at 2.2\,$\mu$m wavelength in air. The measured ratio of yields of the third and fifth harmonics
in our setup is about $2 \cdot 10^{-4}$. This result contradicts the recent suggestion that the Kerr effect 
in air saturates and changes sign in ultra-intense optical fields. 
\end{abstract}


% 320.2250 fs-phenomena
% 320.7110 ultrafast nonlinear optics
% 190.2620 Harmonic generation and mixing
% 190.3270 Kerr effect
% 190.7110 Ultrafast nonlinear optics

 ] %% activate for two-column option

\noindent
Linear and nearly instantaneous dependence of the refractive index of transparent media 
on the intensity of the optical field is the essence of the electronic Kerr effect, which is the cornerstone of
nonlinear optics. In a recent publication, it has been suggested that the notion of the instantaneous electronic Kerr effect 
must be generalized \cite{Loriot}. Measurements of the transient birefringence induced in various
gases by ultraintense and ultrashort optical pulses have been interpreted as the saturation and sign reversal 
of the nonlinear refractive index at the optical intensity levels on the order of several tens of terawatts 
per square centimeter. The common linear dependence of the index of refraction on the intensity 
of the optical field has been amended by the inclusion of new, nontrivially large terms proportional to second and higher 
powers of intensity. 

The field of femtosecond laser filamentation is probably the one that would be affected the most, should the 
above suggestion be proven valid. 
For many years, it was believed that the intrinsically unstable balance between
diffraction and self-focusing in ultraintense laser beams undergoing filamentation in gaseous media is stabilized by
the weak defocusing action of plasma generated on the beam axis via multi-photon ionization
\cite{Mysyrowicz} --\cite{Chin_book}. The higher-order Kerr terms, if they did exist, could provide for an alternative
stabilization mechanism. The filament propagation would be much longer ranged, as the losses to ionization would be 
reduced compared to the established filamentation scenario based on plasma defocusing \cite{WolfPRL}.      

Effects associated with weak saturation of the electronic Kerr response have been considered in the context of 
laser filamentation previously \cite{Akozbek} --\cite{Vincotte}. However, the suggestion that the Kerr effect may reverse sign and 
cause, among other things, beam self-defocusing, were too radical a departure from the established
framework of laser filamentation and nonlinear optics in general. Almost immediately after the publication of 
\cite{Loriot}, the results reported in that paper became the subject of intense debates.

Following \cite{WolfPRL2}, we will refer to the claim put forth in \cite{Loriot} as the Higher-Order Kerr Effect (HOKE) theory.
Several independent experimental tests of HOKE in the context of 
laser filamentation have unambiguously proven that even if the higher-order Kerr terms do exist, they are not operative 
in common laser filaments in gases \cite{Polynkin} --\cite{Milchberg}. The question still remains whether 
this high-order Kerr effect exists at all. It has been argued that the sign reversal of the Kerr effect may
occur in a transient regime, while neutral molecules in the gas are undergoing multi-photon ionization 
\cite{Miro1D},\cite{Kramers}.

Very recent experiments on direct profiling of the cross-phase modulation response in a thin gas target 
illuminated by an ultraintense and ultrashort laser pulse 
revealed no evidence of even a transient saturation of the Kerr effect, in complete contradiction 
with HOKE \cite{MilchbergArXiv}. An alternative test that would also look at the presence of a potentially transient 
HOKE response has been proposed in \cite{MiroOL}. In this Letter, we report experimental results on the realization 
of that proposal. In agreement with \cite{MilchbergArXiv}, we found no evidence of the higher-order Kerr response. 
We believe that at this time, the combined body of experimental evidence is sufficient to permanently disqualify the HOKE paradigm.

The experimental test reported here is based on the dramatic effect that the inclusion of the HOKE
terms would have on the efficiency of third and fifth harmonic generation by a femtosecond laser
pulse with peak power in the vicinity of the alleged turnover point for the higher-order Kerr effect.
As has been numerically shown in \cite{MiroOL} for tightly focused femtosecond pulses at 1.3\,$\mu$m wavelength,
the ratio of the yields of the fifth and third harmonics generated by such pulses is a growing function of the intensity
of the pulse. This ratio saturates at a value on the order of one if the HOKE terms are included in the model, 
and it remains very small ($\sim 10^{-4}$) when these terms are omitted. Such a dramatic difference between
the predictions of the two models is the consequence of the fact that in the established approach, the fifth harmonic generation 
is a cascade process involving the nonlinear mixing between the already generated third harmonic and the leftover
pump, while in the new theory, the fifth harmonic is generated directly from the fundamental through the
higher-order Kerr response.

The above qualitative difference between the predictions of the old and the new models makes the comparison 
of yields of the third and fifth harmonics a reliable test of the new theory that
is largely insensitive to the variations of the material parameters and to a particular experimental 
geometry.        

We point out that an experiment on the generation of third and fifth harmonics in an Argon cell using intense pump
pulses at 800\,nm wavelength has been reported much before the HOKE controversy had even emerged \cite{Old_experiment}. 
The results of that work have been recently used to argue in favor of the validity of HOKE \cite{Kasparian3rdand5th}. 
In our view, \cite{Old_experiment} is not straightforwardly applicable to the resolution 
of the HOKE controversy for the following reasons. Experiments reported in \cite{Old_experiment} were 
certainly not designed to support or disprove the HOKE theory, which was nonexistent at the time  
when \cite{Old_experiment} was published. Instead, that work 
targeted the demonstration of highest possible yield of fifth harmonic. The wavelength of the generated
fifth harmonic was about 162\,nm, which is in vacuum UV, thus making its reliable and quantitative characterization 
not straightforward. 
Furthermore, the extended interaction length and the associated phase mismatch between the pump wavelength
and the generated harmonics,
together with the fact that the actually detected portions of the harmonics were not exactly known, 
made the interpretation of those experiments in the context of the resolution of 
the HOKE controversy not reliable.

The experiment reported here is free from the above complications. 
In our case, third and fifth harmonics are generated using millijoule-level femtosecond
laser pulses at 2.2\,$\mu$m wavelength. This choice of pump wavelength is dictated by three
considerations. First, both third and fifth harmonics of this wavelength fall into the near UV -- visible 
spectral range,
making them easily detectable. Second, the phase-matching distance between the pump beam and
both of the harmonics in air at normal atmospheric pressure, at this pump wavelength, is estimated at over 10 centimeters. 
Thus using a short interaction length of less than 1 centimeter in our experiment 
eliminates any potential issues associated with phase mismatch between the pump and the harmonics.
Third, the effect of the higher-order Kerr terms, according to \cite{Kramers}, should become even more pronounced 
in the mid-infrared, compared to the 800\,nm wavelength range commonly used in the 
high-intensity femtosecond laser experiments.         

Our experimental setup is schematically shown in Figure 1. Femtosecond pulses at 2.2\,$\mu$m
center wavelength are produced by a commercial high-energy Optical Parametric Amplifier (OPA, Palitra-HE model by
Quantronix Corporation). The OPA is pumped by about 40\,fs-long, 20\,mJ pulses at 800\,nm wavelength
generated by an ultrafast Ti:Sapphire amplifier chain operating at a 10\,Hz pulse repetition frequency.
The OPA is tuned to produce signal and idler pulses at 1.26\,$\mu$m and 2.20\,$\mu$m center wavelengths, 
respectively. The output beam from the OPA has a diameter of about 10\,mm. Both signal and idler pulses have
energies of about 1.5\,mJ. The duration of the pulses is 
estimated by the OPA manufacturer at between 40 and 60\,fs \cite{Pancho}.

\begin{figure}[bt]
\begin{center}
\includegraphics[width=8.3cm]{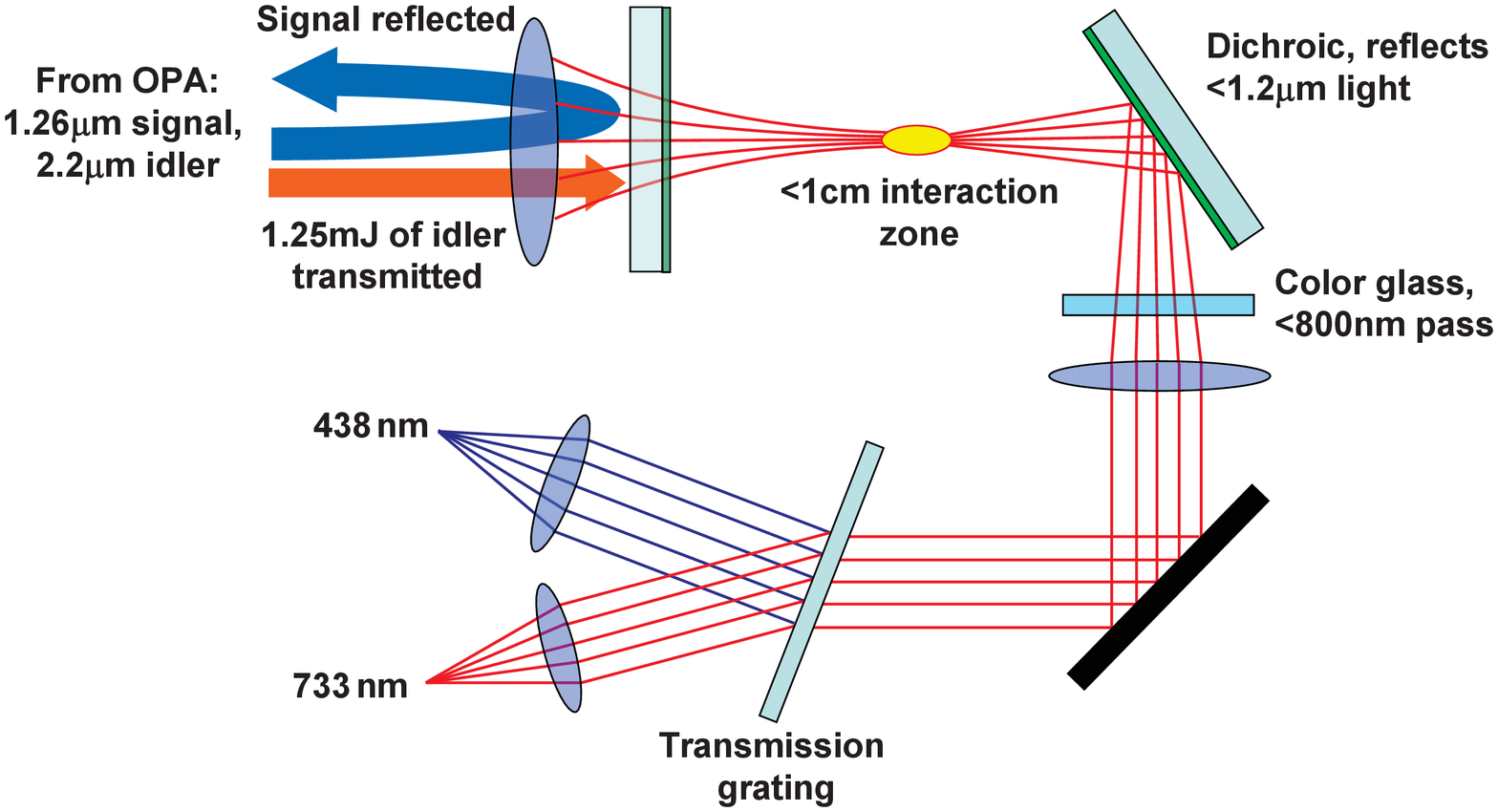}
\caption{(Color online) Schematic of the experimental setup.} 
\label{setup}
\end{center}
\end{figure}

The co-propagating signal and idler beams are focused by a lens with 20\,cm focal length. Immediately after the lens,
the 1.26\,$\mu$m signal is reflected by a dichroic beamsplitter operated at near normal incidence. 
This dichroic (CVI -- Melles Griot Corporation, part number BBDS), is a dielectric low-pass filter.
At normal incidence, it has a transmission cutoff at the wavelength of about 1.3\,$\mu$m. The idler pulse 
at 2.2\,$\mu$m is transmitted by this mirror with about 90\% efficiency, 
while the transmission for the signal beam at 1.26\,$\mu$m is less than
0.1\%. For all wavelengths below 1\,$\mu$m, transmission is less than 0.05\%. 
 
The energy of the focused 2.2\,$\mu$m idler pulse, after the dichroic, is measured at about 1.25\,mJ.
It is important that the dichroic 
is oriented with its substrate facing the incident beam. In this case, the visible and UV light possibly generated inside
the substrate is reflected by the coating, and only the 2.2\,$\mu$m idler pulse is passed into the interaction zone.       

The third and fifth harmonics of the 2.2\,$\mu$m pump are generated in the ambient air, 
near the focal plane of the focusing lens. 
The estimated intensity of the pump beam near the focus is above the 
turnover point for the alleged sign reversal of the Kerr effect in the HOKE model and also above the 
ionization threshold of air. The length of the short and faint plasma spark generated in the vicinity of the interaction zone is
less than 1\,cm. Under these conditions, the ratio of the third and fifth harmonic yields, according
to the HOKE theory, should approach one.

The third and fifth harmonics of the pump are reflected by another dichroic mirror, which is identical to the one 
used to block the signal pulse at 1.26\,$\mu$m from the OPA. The second dichroic is operated with the coating facing the
incident beam and at a 45$^{\circ}$ angle of incidence. This mirror efficiently reflects the generated harmonics, 
but blocks over 90\% of the 2.2\,$\mu$m pump. The residual pump is further attenuated by about five orders 
of magnitude by a 3\,mm\,--thick color-glass filter plate (KG5 filter by Schott Glass Corporation) placed after the
dichroic. 

\begin{figure}[bt]
\begin{center}
\includegraphics[width=8.3cm]{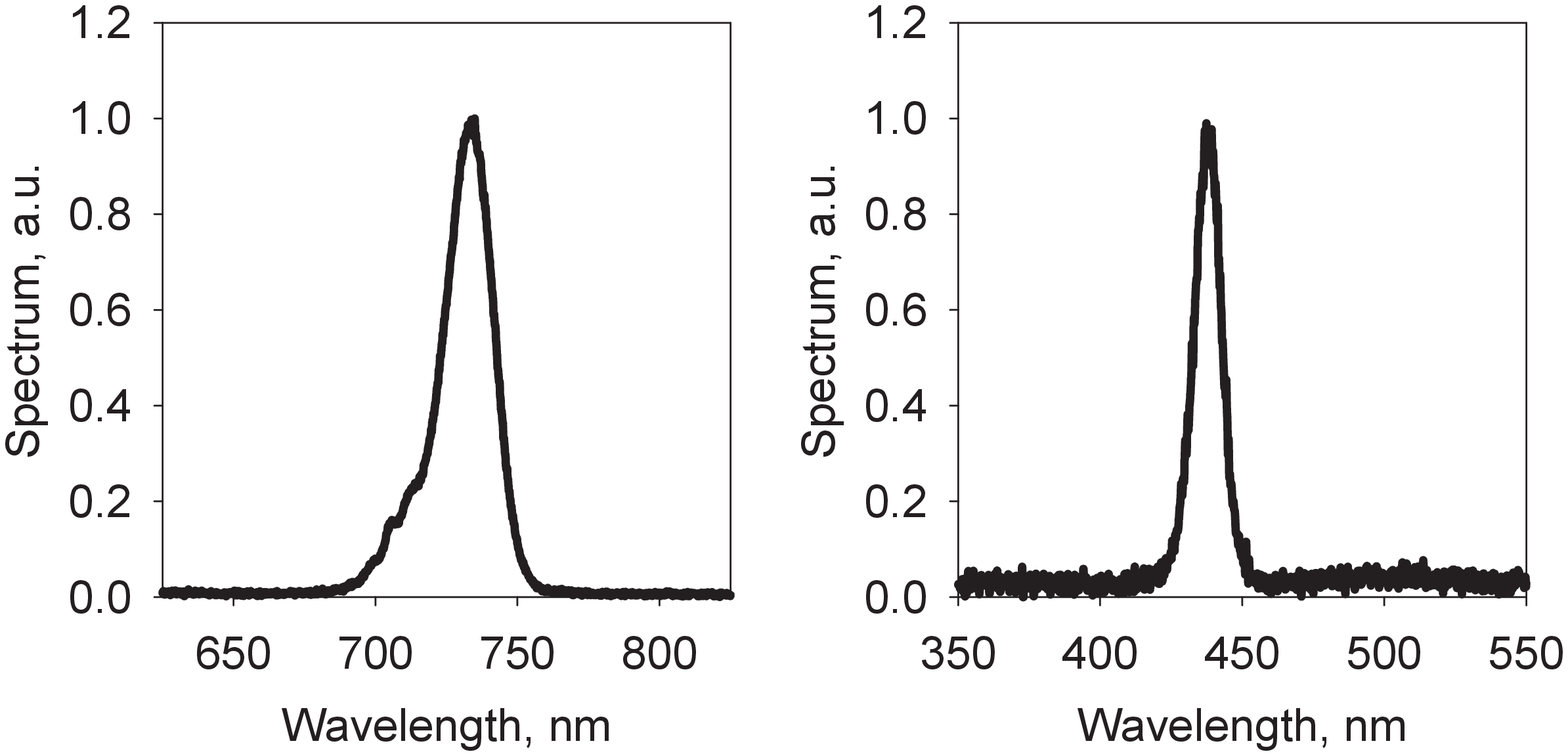}
\caption{Normalized spectra of third harmonic (left) and fifth harmonic (right) generated by 
tightly focused femtosecond pulses at 2.2\,$\mu$m wavelength in air, 
on linear scale.} 
\label{3rd}
\end{center}
\end{figure}

After collimation with another lens, the signals at third and fifth harmonics are spatially 
separated by a 300 lines per millimeter transmissive diffraction grating. 
The spectra of the two harmonics are recorded by 
a fiber-coupled spectrometer (model USB4000 by Ocean Optics Corporation), with the aid of two additional 
lenses that separately focus the harmonic beams after they are separated by the grating. These spectra are shown in 
Figure 2. They are centered at 733\,nm and 438\,nm,
very close to one-third and one-fifth of the pump wavelength of 2.20\,$\mu$m. 

The emission bandwidths of the detected third and fifth harmonics are about 20\,nm and 11\,nm, respectively.
Assuming that the harmonic pulses are both Gaussian and close to transform limit, the durations of the two pulses
can be estimated as 40\,fs and 25\,fs. The ratio of these pulse durations is in the ballpark of 
$\sqrt{5/3}$, which would be expected in the case of the perfectly phase-matched harmonic generation
with non-depleted pump.      

The energies of the two harmonic pulses are measured by dedicated silicon detectors that were  
calibrated against a pyroelectric detector, using reference laser pulses at 800\,nm wavelength. 
Accounting for all losses experienced by the generated harmonics in our setup, 
the energies of the third and fifth harmonics are found equal 
to 1.9\,$\mu$J and 0.4\,nJ, respectively. These pulse energies correspond to yields
of about $1.5 \cdot 10^{-3}$ for third harmonic and $3.0 \cdot 10^{-7}$ for fifth harmonic, with 
respect to the energy of the pump pulse at 2.2\,$\mu$m wavelength in the interaction zone. 
The ratio of yields for the two harmonics is about $2 \cdot 10^{-4}$, in contradiction with the 
HOKE theory.

To verify that the generated harmonics were in fact originating from 
the tightly focused interaction zone
and not from any optics in the setup, we substituted the focusing lens with a glass plate with the same thickness as the lens. 
In that case, the intensities of both third and fifth harmonics of the pump fell below the detectability limit. 

In conclusion, we have conducted a dedicated experiment in order to validate or disprove the existence of the 
higher-order Kerr effect proposed in \cite{Loriot}. Our test was based on the qualitative difference that the 
inclusion of the higher-order Kerr terms has on the yields of the third and fifth harmonics 
generated by tightly focused ultraintense laser pulses at 2.2\,$\mu$m wavelength in ambient air. 
Our results unambiguously disprove the existence of the high-order Kerr response, even in the transient regime. 
Our conclusion is in agreement with another recently reported dedicated test of HOKE \cite{MilchbergArXiv}.      
     
The authors acknowledge helpful discussion with Miroslav Kolesik. This work was supported by 
The United States Air Force Office of Scientific Research under programs 
FA9550-10-1-0237 and FA9550-10-1-0561.

\end{document}